\documentstyle[multicol,fancyheadings,eqsecnum,prd,aps]{revtex}

\begin{document}
\draft 

\title{Gauge transformations in the 
Lagrangian and Hamiltonian formalisms of 
generally covariant theories} 

\author{J. M. Pons\footnote[1]{Electronic address:  
pons@ecm.ub.es}}
\address{Departament d'Estructura i Constituents de la Mat\`eria,
Universitat de Barcelona, \protect\\
Avinguda Diagonal 647, 08028 Barcelona, 
Catalonia, Spain}

\author{D. C. Salisbury\footnote[2]{Electronic address: 
dsalis@austinc.edu}}
\address{Austin College, Sherman, Texas 75090 USA}

\author{L. C. Shepley\footnote[3]{Electronic address: 
larry@helmholtz.ph.utexas.edu}}
\address{Center for Relativity, The University of Texas, Austin, 
Texas 78712-1081, USA}

\date{\today\ --- To appear in Phys.\ Rev.\ D}

\maketitle

\begin{abstract}
We study spacetime diffeomorphisms in Hamiltonian and Lagrangian 
formalisms of generally covariant systems.  We show that the gauge 
group for such a system is characterized by having generators which 
are projectable under the Legendre map.  The gauge group is found to 
be much larger than the original group of spacetime diffeomorphisms, 
since its generators must depend on the lapse function and shift 
vector of the spacetime metric in a given coordinate patch.  Our 
results are generalizations of earlier results by Salisbury and 
Sundermeyer.  They arise in a natural way from using the requirement 
of equivalence between Lagrangian and Hamiltonian formulations of the 
system, and they are new in that the symmetries are realized on the 
full set of phase space variables.  The generators are displayed 
explicitly and are applied to the relativistic string and to general 
relativity.
\end{abstract}

\pacs{04.20.Fy, 04.30.Cv}

 \setlength{\columnseprule}{0pt}\begin{multicols}{2}

\section{Introduction}

The goal of this paper is to understand diffeomorphism symmetries
in the canonical formalism at the classical level.  The putative 
generators of infinitesimal general coordinate transformations feature 
in all canonical quantization approaches, but debate persists in the 
literature as to what aspects of the diffeomorphism group are realized 
at the classical level as canonical transformations 
\cite{Bergmann72,Isham85,Lee90,Salisbury83}.  
This issue is intimately 
related to the meaning of time in quantum gravity.

In this paper we extend recent work by Pons and Shepley \cite{Pons95} 
concerning constrained systems.  We analyze diffeomorphism symmetries 
using in a natural way the equivalence of the Hamiltonian and the 
Lagrangian approaches to generally covariant systems.  We show that 
infinitesimal transformations which are projectable under the Legendre 
map are a basis for the generators of the gauge group.  This group is 
much larger than the original group of spacetime diffeomorphisms 
because it acts on the space of spacetime metrics, whereas the 
diffeomorphism group acts on the underlying manifold.  Since we retain 
the full set of canonical variables, the associated infinitesimal 
generators are new; they are realized on the full set of phase space 
variables and must at least depend in a specific way on the lapse 
function and shift vector of the spacetime metric in a given 
coordinate patch.  The results are contrasted and compared with 
earlier work by Salisbury and Sundermeyer \cite{Salisbury83} on the 
realizeability of general coordinate transformations as canonical 
transformations.

The formalism we shall develop encompasses all generally covariant 
Lagrangian dynamical models containing configuration variables which 
are either metric components or which may be used to construct a 
metric.  We begin in Section 2 with a rederivation of the relation 
between gauge symmetries in Lagrangian and Hamiltonian formalisms.  
After introducing the notions of lapse and shift in Section 3, we show 
that diffeomorphism-induced gauge transformations are projectable 
under the Legendre transformation if and only if infinitesimal 
variations depend on the lapse and shift but not on their time 
derivatives.  These projectable infinitesimal transformations thus 
contain a compulsory dependence on the normal to the chosen time 
foliation.  We illustrate these ideas with the relativistic particle, 
canonical gravity, and the relativistic string.

In Section 4 we turn our attention to the construction of canonical 
generators of the metric-dependent gauge group.  These objects 
generate symmetry transformations on the full set of canonical 
variables.  We show that every generator with non-vanishing time 
component acts as an evolution generator on at least one member of 
every equivalence class of solutions.  Section 5 contains a discussion 
of gauge fixing and the elimination of redundancy in initial 
conditions.  In Section 6, our conclusion, we discuss the nature 
of the diffeomorphism-induced gauge group.  
The Appendix illustrates the projectability conditions in a model, the 
Nambu-Goto string, in which the lapse and shift depend on time 
derivatives of the dynamical variables.


\section{Noether Hamiltonian Symmetries}

We begin by rederiving some results of Batlle {\it et al.} 
\cite{Batlle89} for first order Lagrangians $L(q, \dot q)$.  We 
exclude Lagrangians which explicitly depend on time $t$ since we are 
interested in reparameterization covariant systems.  We start with a 
Noether Lagrangian symmetry,
$$\delta L = {dF}/{dt},
$$ 
and we will investigate the conversion of this symmetry to the 
Hamiltonian formalism.  Defining 
\begin{equation} 
G = ({\partial L}/{\partial \dot q^i}) \delta q^i - F, 
\end{equation} 
we can write 
\begin{equation} [L]_i\delta q^i + \frac{dG}{dt} = 0,
    \label{noet}
\end{equation} 
where $[L]_i$ is the Euler-Lagrange functional derivative of $L$, 
$$ [L]_i = \alpha_i - W_{is}\ddot q^s,
$$
where 
$$ W_{ij}\equiv {\partial^2L\over\partial\dot q^i\partial\dot q^j} 
    \quad {\rm{}and} \quad \alpha_i\equiv 
    - {\partial^2L\over\partial\dot q^i\partial q^s}\dot q^s 
    + {\partial L\over\partial q^i} .
$$

Here we consider the general case where the mass matrix or Hessian 
${\bf{}W}=(W_{ij})$ may be a singular matrix.  In this case there 
exists a kernel for the pullback ${\cal F}\!L^*$ of the Legendre map 
${\cal F}\!L$ from configuration-velocity space $TQ$ (the tangent 
bundle $TQ$ of the configuration space $Q$) to phase space $T^*Q$ (the 
cotangent bundle).  This kernel is spanned by the vector fields
\begin{equation} {\bf\Gamma}_\mu 
    = \gamma^i_\mu {\partial\over\partial\dot q^i} , 
    \label{GAMMA} 
\end{equation} 
where $\gamma^i_\mu$ are a basis for the null vectors of $W_{ij}$.
The Lagrangian time-evolution differential operator can therefore be 
expressed as:
\begin{equation} {\bf X} = {\partial\over\partial t}
    + \dot q^{s}{\partial\over\partial q^{s}}
    +a^{s}(q,\dot q){\partial\over\partial \dot q^{s}}
    +\lambda^{\mu}{\bf\Gamma}_{\mu}
        \equiv {\bf X}_{o} +\lambda^{\mu}{\bf\Gamma}_{\mu} ,
            \label{EVOLOP}
\end{equation} 
where $a^{s}$ are functions which are determined by the formalism, and 
$\lambda^{\mu}$ are arbitrary functions.
It is not necessary to use the Hamiltonian technique to find the 
${\bf\Gamma}_\mu$, but it does facilitate the calculation: 
\begin{equation} \gamma^i_\mu
    = {\cal F}\!L^*\left( \partial\phi_\mu\over\partial p_i \right) , 
    \label{gam} 
\end{equation} 
where the $\phi_\mu$ are the Hamiltonian primary first class 
constraints.

Notice that the highest derivative in (\ref{noet}), $\ddot q^i$, 
appears linearly.  Because $\delta L$ is a symmetry, (\ref{noet}) is 
identically satisfied, and therefore the coefficient of $\ddot q^i$ 
vanishes: 
\begin{equation} 
W_{is} \delta q^s - {\partial G \over\partial\dot q^i} = 0 .  
    \label{w-g} 
\end{equation} 
We contract with a null vector $\gamma^i_\mu$ to find 
that 
$$ {\bf\Gamma}_\mu G = 0 .
$$
It follows that $G$ is projectable to a function 
$G_{\rm H}$ in~$T^*Q$; that is, it is the pullback of 
a function (not necessarily unique) in 
$T^*Q$:
$$
G = {\cal F}\!L^*(G_{\rm H}) .
$$

This important property, valid for any conserved quantity associated 
with a Noether symmetry, was first pointed out by Kamimura 
\cite{kami82}.  Observe that $G_{\rm H}$ is determined up to the 
addition of linear combinations of the primary constraints.  
Substitution of this result in (\ref{w-g}) gives 
$$ W_{is} \left[ \delta q^s - {\cal F}\!L^* 
    \left({\partial G_{\rm H}\over\partial p_s}\right) \right] = 0 ,
$$ 
and so the brackets enclose a null vector of {\bf W}: 
\begin{equation} 
\delta q^i - {\cal F}\!L^*
    \left({\partial G_{\rm H}\over\partial p_i}\right) 
    = \sum_\mu r^\mu \gamma^i_\mu ,
                                      \label{dq}
\end{equation} 
for some $r^\mu(t, q, \dot q)$.

We shall investigate the projectability of variations generated by 
diffeomorphisms in the following section.  Assume for now that an 
infinitesimal transformation $\delta q^i$ is projectable: 
$${\bf\Gamma}_\mu \delta q^i =0.
$$
Notice that if $\delta q^i$ is projectable, so must be $r^\mu$, 
so that $r^\mu = {\cal F}\!L^* (r^\mu_{\rm H})$.  
Then, using (\ref{gam}) and (\ref{dq}), we see that
$$\delta q^i 
  = {\cal F}\!L^*
  \left({\partial (G_H + \sum_\mu r^\mu_{\rm H}\phi_\mu)\over\partial  
  {p_i}}\right).
$$
We now redefine $G_{\rm H}$ to absorb the piece 
$\sum_\mu r^\mu_{\rm H} \phi_\mu$, and from now on we will have
$$  \delta q^i = {\cal F}\!L^*
\left({\partial G_{\rm H} \over\partial p_i}\right).  
$$

Define
$$  \hat p_i = {\partial L\over\partial\dot q^i}; 
$$
after eliminating (\ref{w-g}) times $\ddot q^i$ from (\ref{noet}), 
we obtain 
\begin{eqnarray}
     \left({\partial L\over\partial q^i} 
    - \dot q^s {\partial \hat p_i\over\partial q^s} \right) 
    {\cal F}\!L^*({\partial G_{\rm H}\over\partial p_i})
    &+& \dot q^i {\partial \over\partial q^{i}}
            {\cal F}\!L^* (G_{\rm H})  \nonumber\\
    &+& {\cal F}\!L^* ({\partial G_{\rm H}\over\partial t}) =0,
\end{eqnarray}
which simplifies to 
\begin{equation} 
{\partial L\over\partial q^i} {\cal F}\!L^*( 
{\partial G_{\rm H}\over\partial p_i} )  
    + \dot q^i {\cal F}\!L^*({\partial G_{\rm H}\over\partial q^i}) 
    + {\cal F}\!L^* 
    ({\partial G_{\rm H}\over\partial t}) =0.
        \label{noet-wg}
\end{equation}
Now let us invoke two identities \cite{Batlle86} that are at the core 
of the connection between the Lagrangian and the Hamiltonian 
equations of motion.  They are
$$  \dot q^i = {\cal F}\!L^* ({\partial H\over\partial p_i}) 
    + v^\mu(q, \dot q) {\cal F}\!L^*
    ({\partial \phi_\mu\over\partial p_i}), 
$$ 
and
$$  {\partial L\over\partial q^i} 
= - {\cal F}\!L^* ({\partial H\over\partial q^i}) 
    - v^\mu(q, \dot q) {\cal F}\!L^*
    ({\partial \phi_\mu\over\partial q^i}); 
$$ 
where $H$ is any canonical Hamiltonian, so that 
${\cal F}\!L^*(H) 
= \dot q^i (\partial L / \partial \dot q^i) - L =\hat E$, 
the Lagrangian energy, and the functions $v^\mu$ are 
determined so as to render the first relation an identity.  
Notice the important relation 
\begin{equation} {\bf\Gamma}_\mu v^\nu = \delta_\mu^\nu, 
\end{equation} 
which stems from applying ${\bf\Gamma}_\mu$ to the first identity 
and taking into account that 
\[ {\bf\Gamma}_\mu \circ{\cal F}\!L^* = 0. 
\]
Substitution of these two identities into (\ref{noet-wg}) yields
(where $\{\,,\,\}$ is the Poisson Bracket)
$$ {\cal F}\!L^*\{G_{\rm H},H\} 
        + v^\mu {\cal F}\!L^*\{G_{\rm H},\phi_\mu \} 
        +{\cal F}\!L^* ({\partial G_{\rm H}\over\partial t}) 
     =0.
$$
This result can be split through the action of ${\bf\Gamma}_\mu$ into 
\begin{equation} 
    {\cal F}\!L^*\{G_{\rm H},H\} 
    +{\cal F}\!L^* ({\partial G_{\rm H}\over\partial t}) =0, 
        \nonumber 
\end{equation} 
and 
\begin{equation} {\cal F}\!L^*\{G_{\rm H},\phi_\mu \} = 0; \nonumber 
\end{equation} 
or equivalently, 
\begin{equation} \{G_{\rm H},H\} 
    + ({\partial G_{\rm H}\over\partial t}) = pc ,
    \label{dgdt}
\end{equation} 
and 
\begin{equation} \{G_{\rm H},\phi_\mu \} = pc,
    \label{g,phi}
\end{equation} 
where $pc$ stands for any linear combination of primary 
constraints.  We have arrived at a neat characterization for a 
generator $G_{\rm H}
$ of Noether transformations in the canonical formalism.

Up to now we have considered general Noether symmetries, encompassing 
rigid (global) as well as gauge (local) transformations.  Let us 
finally specialize to gauge transformations.  For reparameterization 
covariant theories, except for a small number of exceptional cases not 
important for this paper \cite{pep94}, 
a gauge generator will be of the form 
$$ G_{\rm H}(t) = \epsilon(t) G_0(q,p) + \dot\epsilon(t)G_1(q,p), 
$$ 
where $\epsilon(t)$ is an arbitrary function.  
Because of the arbitrariness 
of $\epsilon(t)$, and recognizing that the Poisson Bracket of the 
Hamiltonian with primary constraints yields secondary constraints, we 
learn from (\ref{dgdt}) that
$$
G_1 = pc,
$$
\begin{equation} G_0 = - \{G_1,H\} + pc,
    \label{ggen}
\end{equation} 
and 
\begin{equation} \{G_0,H\} = pc;
    \label{ggen2}
\end{equation} 
while from (\ref{g,phi}) we deduce that 
\begin{equation} \{G_0, pc\} = pc, \label{test1}
\end{equation}
and
\begin{equation} \{G_1, pc\} = pc. 
    \label{test2} 
\end{equation} 
It can be shown from 
(\ref{ggen}) that $G_0$ must contain a piece which is a secondary 
constraint, while (\ref{test1}) and (\ref{test2}) show that 
both $G_0$ and the primary 
constraint $G_1$ are first class.


\section{Diffeomorphism-Induced Gauge Symmetries}

We specialize now to generally covariant dynamical models in which a 
metric can be constructed with the configuration variables (but not 
with velocity variables).  We assume in addition that no further gauge 
symmetry exists.  We shall illustrate our results with the 
relativistic particle with an auxiliary variable and with general 
relativity.  Our first objective is to determine the general form of 
projectable variations resulting from diffeomorphisms on a coordinate 
patch.

If a metric exists in a coordinate system {$\{x^\mu\}$} the line 
element may always be written in the form 
\begin{equation} 
    ds^2=-N^2 (dx^0)^2 + g_{ab}(N^adx^0+dx^a)(N^bdx^0+dx^b) 
        \label{3.1} 
\end{equation} 
with contravariant metric components given by 
\begin{equation} 
(g^{\mu \nu}) = \pmatrix{ 
          -N^{-2}   & N^{-2}N^a   \cr 
          N^{-2}N^a & e^{a b}-N^{-2}N^a N^b
          } , \label{3.2}
          \end{equation} 
with $e^{ab}g_{bc}=\delta^{a}_{c}$.
The lapse function $N$ and shift vector $N^a$ will play important 
roles in our discussion.  Our index conventions are 
that greek indices range from 0 to $M$, where $M$ is the 
dimension of the spacelike hypersurfaces of the time foliation.  
Latin indices range from 1 to $M$.

Explicitly, the configuration space variables are $N^\mu$ (with 
$N^0 \equiv N$) and $g_{a b}$.  The unit normal $n^{\mu}$ to 
the spacelike hypersurfaces is given by 
\begin{equation} 
    n^{\mu}=\delta^{\mu}_0 N^{-1} - \delta^{\mu}_a N^{-1} N^a\ 
    {\rm{},\ so\ that}\ n^\mu n^\nu g_{\mu\nu}=-1.  
        \label{3.3} 
\end{equation} 
Since $e^{a b}$ is the inverse of the three-metric $g_{a b}$, the 
contravariant components of the spacetime metric are
\begin{equation} 
    g^{\mu \nu}=e^{ab}\delta^\mu_a\delta^\nu_b - n^\mu n^\nu.  
    \label{3.4} 
\end{equation} 
Diffeomorphism covariance prevents the 
lapse $N$ and shift $N^a$ from being fixed by the equations 
of motion in any generally covariant dynamical model.  
Specifically, since the $N^{\mu}$ are arbitrary, $\ddot N^{\mu}$
are undetermined.  The evolution operator (\ref{EVOLOP}) acting on
$\dot N^{\mu}$ must therefore serve only to relate the arbitrary 
functions $\lambda^{\mu}$ to the $\ddot N^{\mu}$.  
Consequently, ${\bf \Gamma}_{\mu}\dot N^{\nu}$ must 
form a  nonsingular matrix.  
Further, ${\bf \Gamma}_{\mu}$ acting on any other 
velocity must give zero, since we are assuming no other 
gauge symmetry.  It follows that the null 
vectors of the Hessian {\bf W} (see \ref{GAMMA}) are spanned by
\begin{equation} 
    {\bf{\Gamma}}_{\mu}={\partial \over \partial {\dot N }^{\mu}}.  
     \label{3.5} 
\end{equation} 
Since there are $M+1$ of the $N^\mu$, these null vectors span the 
arbitrary component of the Lagrangian evolution operator 
(\ref{EVOLOP}).

Now consider infinitesimal coordinate transformations $x^\mu 
\rightarrow x^\mu - \epsilon^\mu(x)$ , with $\epsilon^\mu$ arbitrary 
functions of the coordinate variables $x^\nu$.  The corresponding 
variations of the components of the metric tensor (the Lie 
derivative of the metric along $\epsilon^\mu$) are 
(${}_{,\mu}\equiv\partial/\partial x^\mu$):
\begin{equation}
    \delta g^{\mu\nu} 
        = g^{\mu\nu}_{~~,\rho} \epsilon^\rho 
        - g^{\mu\rho} \epsilon^\nu_{,\rho} 
        - g^{\rho\nu} \epsilon^\mu_{,\rho},  
            \label{3.6} 
\end{equation} 
or 
\begin{equation} 
    \delta g_{\mu\nu} = g_{\mu\nu,\rho} \epsilon^\rho 
        + g_{\mu\rho} \epsilon^\rho_{,\nu} 
        + g_{\rho\nu} \epsilon^\rho_{,\mu}.  
            \label{3.6a} 
\end{equation} 
The variations of the $N^\mu$ are readily calculated: 
\begin{equation} 
    \delta N = N_{,\mu} \epsilon^\mu 
        +N \epsilon^0_{,0} -N N^a \epsilon^0_{,a}, 
    \label{3.8} 
\end{equation}
\begin{eqnarray} 
    \delta N^a &=& N^a_{,\mu} \epsilon^\mu 
        + N^a \epsilon^0_{,0} -(N^2 e^{a b} 
        + N^a N^b)\epsilon^0_{,b} \nonumber\\
        &&~~~~~~+\epsilon^a_{,0} -N^b \epsilon^a_{,b}.  
            \label{3.9} 
\end{eqnarray} 
Thus the variations of the $N^\mu$ do depend on ${\dot N}^\mu = 
N^\mu_{,0}$ (but the variations of $g_{a b}$ do not), assuming as we 
have above, that $\epsilon^\mu$ depends only on the coordinates.  
Consequently the variations of $N^\mu$ are clearly not projectable; 
projectability is attained only if we permit $\epsilon^\mu$ to depend 
on $N^\mu$.  The requirement that derivatives of $\delta N^\mu$ with 
respect to ${\dot N}^\mu$ vanish implies that
\begin{equation} 
    \epsilon^0 + N {\partial \epsilon^0 \over \partial N} = 0, 
    \label{3.10} 
\end{equation} 
\begin{equation} {\partial \epsilon^0 \over \partial N^a }= 0, 
    \label{3.11} 
\end{equation} 
\begin{equation} N^a {\partial \epsilon^0 \over \partial N} 
    + {\partial \epsilon^a \over \partial N} = 0, 
        \label{3.12} 
\end{equation}
\begin{equation} 
    \epsilon^0 \delta^a_b 
        + {\partial \epsilon^a \over \partial N^b}=0.  
    \label{3.13} 
\end{equation} 

These equations were first obtained in \cite{Salisbury83} using a 
rather different approach.  In \cite{Salisbury83}, the following 
requirement was introduced for diffeomorphism-induced gauge 
transformations: Consider $\delta_1 x^\mu = -\epsilon_1^\mu(x, g(x))$ 
and $\delta_2 x^\mu = -\epsilon_2^\mu(x, g(x))$; then ask for 
conditions to be satisfied by $\epsilon_1, \epsilon_2$ such that 
$[\delta_1,\delta_2] x^\mu $ has no explicit time derivatives of 
$\epsilon_1$ or $\epsilon_2$.  We will discuss in the next section the 
reason why this latter approach gives results coincident with ours.  
We feel that the requirement of projectability (independence of the 
$\delta N^\mu$ on ${\dot N}^\mu$ in this case) is a more natural 
approach.

The general solution of the $\epsilon^\mu$ equations 
(\ref{3.10}-\ref{3.13}) is 
\begin{equation} \epsilon^\mu = \delta^\mu_a \xi^a + n^\mu \xi^0,\ 
    {\rm{}so\ that\ } 
    \epsilon^0={\xi^0\over N},\ 
    \epsilon^a=\xi^a - {N^a\over N} \xi^0, 
        \label{3.14} 
\end{equation} 
where $\xi^a $ and $\xi^0$ are arbitrary functions of the spacetime 
variables $x^\mu$ and $g_{ab}$ but are independent of $N^\mu$.  The 
dependence on the $M$-surface metric plays no role in our present 
arguments but is required, as we show in Section 6, in order that the 
diffeomorphism-induced transformations form a group.  The result 
(\ref{3.14}) is true in a more general context than we have been 
treating.  The Appendix will illustrate this point with an example, 
the Nambu-Goto string, in which the metric is built with velocity 
variables as well as configuration space variables.


\subsection{The Free Relativistic Particle with Auxiliary Variable}

We illustrate first with the unit-mass relativistic free particle 
model with auxiliary variable described by the Lagrangian 
\begin{equation}  L = \frac{1}{2e} \dot x^\mu \dot x^\nu \eta_{\mu\nu} 
    - \frac{1}{2} e, 
        \label{partlag}
\end{equation} 
where ${ x^{\mu}}$ is the vector variable in Minkowski spacetime, 
with metric $(\eta_{\mu\nu})={\rm{}diag}(-1,1,1,1)$, and $e$ is an 
auxiliary variable whose equation of motion gives 
$e = (-{\dot x}^\mu {\dot x}_\mu)^{1/2}$.  
Substituting this value of $e$ into the 
Lagrangian leads to the free particle Lagrangian 
$-(-{\dot x}^\mu {\dot x}_\mu)^{1/2}$.

The following Noether gauge transformation is well-known to describe
the reparameterization invariance for this Lagrangian 
($\delta L = {d\over dt}(\epsilon L)$): 
\begin{equation} 
    \delta x^\mu = \epsilon {\dot x}_{\mu}, \ \delta e 
    = \epsilon \dot e + {\dot \epsilon} e.  
    \label{3.16} 
\end{equation} 
Here $\epsilon$ is an infinitesimal arbitrary function of the 
evolution parameter $t$.  Comparing (\ref{3.16}) with (\ref{3.8}), 
we observe that $e$ may be interpreted as a lapse, with 
corresponding metric $g_{0 0}=-e^2$.

The kernel of the pullback map ${\cal F}\!L^*$ is defined in 
(\ref{GAMMA}); here it is spanned by the vector field 
${\bf\Gamma} = \partial/ \partial \dot e$.  
The condition that a function $f$ in 
configuration-velocity space be projectable to phase space is
$$  {\bf\Gamma} f = \frac{\partial f}{\partial \dot e} = 0.  
$$ 
The Noether transformation (\ref{3.16}) is not projectable to phase 
space, since ${\bf\Gamma} \delta e \neq 0$.  Projectable 
transformations are of the form (\ref{3.14}): 
\begin{equation} 
    \epsilon(t,e) = {\xi(t)}/{e}.  
    \label{proj3.16} 
\end{equation} 
The Noether variations then become: 
\begin{equation} 
    \delta x^{\mu} = \xi \frac{{\dot x}^\mu}{e},\ \delta e = \dot \xi.  
    \label{del-x} 
\end{equation}

The arbitrary function describing the Noether gauge transformation is 
$\xi(t)$.  What we have achieved is a change of the generator of the 
gauge transformations.  This leads to a change of the gauge algebra 
which in our case becomes Abelian.  But from the point of view of the 
gauge symmetry of our model we still have the same mappings of 
solutions onto gauge equivalent solutions.  That is, on a given 
dynamical trajectory $ x^\mu_0(t), \, e_0(t)$ we can match the 
transformation given by an arbitrary $\epsilon_0(t)$ with $\xi_0(t)$ 
defined as $\xi_0(t) \equiv e_0(t) \epsilon_0(t)$.  It is in comparing
a transformation on one dynamical trajectory with that acting on 
another where the change has occurred.  In Section 6 we shall 
elaborate further on the issue of the gauge group.

The canonical Hamiltonian is
$$  H = \frac{1}{2} e ( p^\mu p_\mu + 1),
$$
and there is a primary constraint $\pi \simeq 0$, where $\pi$ is the 
variable conjugate to $e$.  The evolution operator vector field 
$\{-,H\}+ \lambda(t) \{-,\pi\}$ yields the secondary constraint 
$\frac{1}{2} (p^\mu p_\mu + 1) \simeq 0$.  Both the primary and the 
secondary constraints are first class.  The arbitrary function 
$\lambda$ is a reflection of the gauge invariance of the model.  The 
solutions of the equations of motion are:
$$  x^\mu(t) = x^\mu(0) + p^\mu(0) \left(e(0) t 
    + \int_0^t d\tau \int_0^\tau d\tau' \, \lambda(\tau')\right), 
$$
$$  e(t) = e(0) + \int_0^t d\tau \, \lambda(\tau),
$$
$$  p^\mu(t) = p^\mu(0),
$$
$$  \pi(t) = \pi(0),
$$
with the initial conditions satisfying the constraints.

Gauge transformations relate trajectories obtained through different 
choices of $\lambda(t)$.  Consider an infinitesimal change 
$\lambda \rightarrow \lambda + \delta\lambda$.  Then the change 
in the trajectories (keeping the initial conditions intact) is:
$$  \delta x^\mu(t) = p^\mu(0) 
    \left(\int_0^t d\tau \int_0^\tau d\tau' \, 
    \delta\lambda(\tau')\right), 
$$
$$  \delta e(t) = \int_0^t d\tau \, \delta\lambda(\tau),
$$
$$  \delta p^\mu(t) = 0 , \, \delta \pi(t) = 0,
$$
which is nothing but a particular case of the projectable gauge 
transformations displayed above with 
$$\xi(t) =\int_0^t d\tau \int_0^\tau d\tau' \, \delta\lambda(\tau').
$$


\subsection{Diffeomorphisms in Canonical General Relativity}

Up to a boundary piece, the Einstein-Hilbert Lagrangian can be written 
as \cite{Wald84} 
\begin{equation} 
    {\cal L} = ({}^{3}\!g)^{\frac{1}{2}} N ({}^3\!R + K_{ab}K^{ab} 
           - K^2) , \label{lagr} 
\end{equation} 
where ${}^{3}\!g$ is the determinant of the 3-metric tensor in 
(\ref{3.1}), ${}^{3}\!g = \det(g_{ab})$, ${}^3\!R$ is the scalar 
curvature computed from the 3-metric, and $K_{ab}$ is the second 
fundamental form (extrinsic curvature) for the constant-time 
3-surfaces :
\begin{equation} 
    K_{ab}={1\over 2N}(\dot g_{ab} - N_{a|b} - N_{b|a}), 
\end{equation}
with ${}_|$ meaning covariant differentiation with respect to the 
3-metric connection.  Notice that the lapse $N$ and shift $N^a$ of the 
4-metric all appear, but their time-derivatives do not.

We may directly apply our general formalism, with no notational 
changes, to conclude that projectable infinitesimal coordinate 
transformations must be of the form 
\begin{equation} 
    x^\mu \rightarrow x^\mu - \delta^\mu_a \xi^a - n^\mu \xi^0.  
    \label{sptm3.xx} 
\end{equation} 
Notice also that for any specific spacetime metric $g_{\mu\nu}(x)$ we 
can implement {\it any} infinitesimal diffeomorphism $x^\mu 
\rightarrow x^\mu - \epsilon^\mu(x)$ by taking the set $\xi^0, \xi^i$ 
(assuming $N \neq 0$) as:
$$
    \xi^0 = N \epsilon^0 , \quad \xi^a = \epsilon^a + N^a \epsilon^0; 
$$
therefore we are not restricting the (infinitesimal) diffeomorphisms 
that can act on any specific metric.  What we achieved is a set of 
generators of the gauge group which can be projected to the phase 
space.


\section{The Hamiltonian Gauge Generators}

Our objective in this section is to derive the full set of 
diffeomorphism-induced gauge generators for the class of dynamical 
models treated in Section 2.  Since the $\xi^{\mu}$ are now the 
arbitrary functions of time appearing in the variations $\delta q^i$ 
of Section 2, we modify the argument leading to the general form of 
the symmetry generators to conclude that these generators must be of 
the form
$$  G(t) = \int d^M\!x \left(\xi^\mu({\bf x},t) G^{(0)}_\mu 
    + \dot\xi^\mu({\bf x},t)G^{(1)}_\mu\right).  
$$

When we use $\epsilon^\mu = \delta^\mu_a \xi^a + n^\mu \xi^0$ in 
(\ref{3.6}), the algebra of the infinitesimal transformations ceases 
to be the standard diffeomorphism algebra.  The standard algebra is 
that of Lie derivatives: 
$\epsilon^\mu_3 = \epsilon^\nu_2 \epsilon^\mu_{1,\nu} 
    - \epsilon^\nu_1 \epsilon^\mu_{2,\nu} 
    =({\cal L}_{\epsilon_2}\epsilon_1)^{\mu} $.  
In our case the commutator of two infinitesimal transformations yields 
an $\epsilon^\mu_3$ of the form of (\ref{3.14}), with 
\begin{equation} 
    \xi_3^a = \xi^b_2\xi^a_{1,b} - \xi^b_1\xi^a_{2,b} 
    - e^{a b}\xi^0_1 \xi^0_{2,b} 
    + e^{a b}\xi^0_2 \xi^0_{1,b}, 
        \label{comrel-a}
\end{equation} 
\begin{equation} 
    \xi_3^0 = \xi^a_2 \xi^0_{1,a} - \xi^a_1 \xi^0_{2,a} .  
    \label{comrel-0}
\end{equation} 

These are the new commutation relations of the gauge algebra in 
configuration-velocity space.  The commutation rules of the gauge 
generators in phase space coincide with the commutation relations in 
configuration-velocity space as long as at least all but one of the 
$M+1$ gauge generators are linear in the momenta (see 
\cite{Batlle89a}).  This amount of linearity holds in our models: 
(\ref{3.14}) implies that time-independent $M$-space diffeomorphisms 
are always projectable.  This means that each associated canonical 
generator must be linear in the momenta; otherwise the transformation 
of configuration variables will depend on velocities.  We can 
conclude that we know the Poisson Bracket rules 
$G[\xi_3] = \{G[\xi_1], G[\xi_2] \}$ 
for our gauge generators.  Thus in comparing 
(\ref{comrel-a}) and (\ref{comrel-0}) with $G[\xi_3]$ we deduce the 
algebra for these generators.  In what follows, we use the convention 
that repeated indices imply both summation and M-dimensional 
integration; we use primed indices where necessary to make sure that 
separate integrations are clearly delineated, though we drop the 
primes on the indices where no loss of clarity is involved:
\begin{equation} 
    \{G_\mu^{(0)},G_{\nu '}^{(0)}\}
    =C_{\mu \nu '}^{\alpha ''} G_{\alpha ''}^{(0)} 
    +({d \over dt}C_{\mu {\nu '}}^{\alpha ''} ) G_{\alpha ''}^{(1)}, 
            \label{5.20} 
\end{equation} 
\begin{equation} 
    \{G_\mu^{(0)},G_{\nu '}^{(1)}\}
    =C_{\mu \nu '}^{\alpha ''} G_{\alpha ''}^{(1)}, 
        \label{5.21} 
\end{equation}
and 
\begin{equation} 
    \{G_\mu^{(1)},G_{\nu '}^{(1)}\}=0, 
    \label{5.22} 
\end{equation}
where the structure coefficients are given by 
\begin{eqnarray} 
    C_{0 0'}^{a''}
    &=& e^{a b}({\bf x}'')\bigg(\delta^M ({\bf x}-{\bf x}'')
       \nonumber\\
        &&~~~~+\delta^M ({\bf x}'-{\bf x}'')\bigg) 
         {\partial \over \partial x^b} \delta^M ({\bf x}-{\bf x}'), 
                \label{5.22a} 
\end{eqnarray} 
\begin{equation} C_{0 0'}^{0''}=0, \label{5.22b} 
\end{equation} 
\begin{equation} C_{a 0'}^{0''}
        =\delta^M ({\bf x}-{\bf x}'') 
            {\partial \over \partial x^a} \delta^M ({\bf x}-{\bf x}') 
        = - C_{0' a}^{0''}, \label{5.22c} 
\end{equation} 
\begin{equation} C_{a 0'}^{b''}= 0, \label{5.22d} 
\end{equation} 
\begin{eqnarray} C_{a b'}^{c''}&=&
        \bigg(\delta_a^c \delta^M ({\bf x}''-{\bf x}')
            {\partial \over \partial x^b}  \nonumber\\
        &&~~~~+ \delta_b^c \delta^M ({\bf x}''-{\bf x})
            {\partial \over \partial x^a}\bigg) 
            \delta^M ({\bf x}-{\bf x}'), \label{5.22e} 
\end{eqnarray} 
and 
\begin{equation} C_{a b'}^{0''}=0.  \label{5.22f} 
\end{equation}

We now construct these generators explicitly.  The canonical 
Hamiltonian (such that its pullback under the Legendre transformation 
gives the Lagrangian energy) for the class of models under discussion 
is 
\begin{equation} 
    H = N^\mu {\cal H}_\mu,         \label{444}
\end{equation} 
where the ${\cal H}_\mu$ are independent of $N^\mu$ and $P_\mu$, and 
the primary constraints are $P_\mu$, the canonical variables conjugate 
to $N^\mu$.  The secondary constraints are 
$\dot P_\mu = \{P_\mu , H \} = - {\cal H}_\mu$, 
and no more constraints appear.  It was shown in 
\cite{Pons95} that the canonical Hamiltonian always takes the form 
(\ref{444}).  All constraints are required to be first class; the 
reason is that $\xi^\mu$ and $\dot\xi^\mu$ appear in (\ref{3.6}), and 
so all constraints are required to build the spacetime gauge 
generators that our theory possesses.

The Dirac Hamiltonian $H_{\rm D}$ is constructed by the addition to 
$H$ of a linear combination (with arbitrary functions 
$\lambda^\mu$) of the primary constraints: 
\begin{equation} 
    H_{\rm D} = H + \lambda^\mu P_\mu.  
    \label{dirach} 
\end{equation} 
$G^{(1)}_\mu$ must be a primary constraint, so the simplest choice is 
$G^{(1)}_\mu = P_\mu$.  It is now necessary to apply (\ref{ggen}): 
$ G^{(0)}_\mu = -\{G^{(1)}_\mu , H \} + pc, $ implying
$$
    G^{(0)}_\mu = {\cal H}_\mu + A_\mu^\nu P_\nu .
$$ 
From (\ref{5.20}) we deduce that
$$
  \{{\cal H}_\mu, {\cal H}_\nu \} 
  = C^\sigma_{\mu\nu}{\cal H}_\sigma ,
$$ 
since $\{N^\mu,{\cal H}_\nu \}=\{P_\mu,{\cal H}_\nu \}=0$.

The $A^\nu_\mu$ are determined by applying condition (\ref{ggen2}) to 
$G^{(0)}_\mu$:
\begin{eqnarray} 
    \nonumber pc &=& \{{\cal H}_\mu + A_\mu^\nu P_\nu , H \} \\ 
     \nonumber &=&  N^\nu \{{\cal H}_\mu, {\cal H}_\nu \} 
         + A^\nu_\mu \{P_\nu , H \} \\
     \nonumber &=& N^\nu C^\sigma_{\mu\nu}{\cal H}_\sigma - A^\nu_\mu 
         {\cal H}_\nu, 
\end{eqnarray} 
which implies
$$  A_\mu^\nu = N^\rho C^\nu_{\mu\rho}
$$
up to an irrelevant arbitrary linear combination of primary 
constraints that would add an ineffective piece to the gauge 
generator.  (By ineffective we mean that the added piece is quadratic 
in the constraints.)  We ignore this piece and take the 
simplest solutions available for $A_{\mu}^{\nu}$.

It is trivial to check the fulfillment of conditions 
(\ref{test1},\ref{test2}).  By use of the Jacobi identity we find
\begin{eqnarray} \nonumber 
  0 &\equiv& \{{\cal H}_\alpha,\{{\cal H}_\beta,{\cal H}_\gamma \}\}
    + \{{\cal H}_\beta,\{{\cal H}_\gamma,{\cal H}_\alpha \}\} \\
    &~~~~&+ \{{\cal H}_\gamma,\{{\cal H}_\alpha,{\cal H}_\beta \}\}
        \nonumber  \\
  \nonumber &=&\left(C_{\beta \gamma}^\rho C_{\alpha \rho}^\sigma 
    + C_{\gamma \alpha}^\rho C_{\beta \rho}^\sigma 
    + C_{\alpha \beta}^\rho C_{\gamma \rho}^\sigma\right) 
         {\cal H}_\sigma 
      + \{{\cal H}_\alpha,C_{\beta \gamma}^\rho \} {\cal H}_\rho\\ 
  &~~~~&  
    + \{{\cal H}_\beta,C_{\gamma \alpha}^\rho \} {\cal H}_\rho 
    + \{{\cal H}_\gamma,C_{\alpha \beta}^\rho \} {\cal H}_\rho, 
        \label{5.22g} 
\end{eqnarray} 
together with 
\begin{equation} {d \over dt} C_{\alpha \beta}^\gamma 
    = N^\rho \{C_{\alpha \beta}^\gamma, {\cal H}_\rho \}, 
        \label{5.22h} 
\end{equation} 
and it is straightforward to show that 
the generators $G_\mu^{(0)}$ and $G_\mu^{(1)}$ do satisfy the algebra 
(\ref{5.20}-\ref{5.22}).

We have therefore obtained the full set of 
diffeo\-mor\-phism-induced gauge generators: 
\begin{equation} 
    G(t) = P_\mu \dot\xi^\mu 
    + ( {\cal H}_\mu + N^\rho C^\nu_{\mu\rho} P_\nu) \xi^\mu,
         \label{thegen}
\end{equation} 
(where the repeated index, to repeat, involves an integration).  
Note that $G(t)$ generates variations in the full phase space.  It is 
straightforward to verify that it does generate the correct 
variations of $N^{\mu}$ (\ref{3.8}-\ref{3.9}) under the 
diffeomorphism-induced gauge transformations (\ref{3.14}).

The preceding discussion applies with no modification of notation to 
canonical general relativity.

We continue with some general remarks on diffeomorphism generators.  
Our first observation is that every generator $G(t)$ with 
$\xi^0 \neq 0$ is interpretable as a global 
time translation generator for at 
least one member of every gauge equivalence class of solutions.  To 
demonstrate this property we note that for a given set of functions 
$\xi^\mu$ in the expansion
\begin{equation} 
    \epsilon^\mu = \delta^\mu_a \xi^a + n^\mu \xi^0,  
    \label{diff4.xx} 
\end{equation} 
we can solve for the $N^\mu$ which render 
$\epsilon^\mu = \delta^\mu_0$: 
\begin{equation} 
    \epsilon^\mu = \delta^\mu_0 
    =\delta^\mu_a \xi^a 
    + (N^{-1}\delta^\mu_0-N^{-1}N^a \delta^\mu_a)\xi^0.  
        \label{solv.xx} 
\end{equation} 
The solution is $N^\mu = \xi^\mu$, which renders $G(t)$ in 
(\ref{thegen}) identical to the Dirac Hamiltonian (\ref{dirach}), 
once we take into account that the equations of motion provide 
$\dot{N}^\mu = \lambda^\mu$.  Therefore the gauge 
generator contains, for any solution of the equations of motion, 
the dynamical evolution as a particular case.

At this point we are ready to understand the coincidence of the two 
approaches mentioned in the previous section: The same conditions 
(\ref{3.10}-\ref{3.13}) are obtained if (1) one asks for the 
projectability of (\ref{3.6a}), or (2) one asks for 
$[\delta_1,\delta_2] x^\mu $ not to have any explicit time 
derivative of $\epsilon_1$ or $\epsilon_2$.  It seems odd that 
conditions imposed on {\it one} transformation (projectability) 
and conditions imposed on the commutation of {\it two} 
transformations should give the same results.  
The reason lies in the structure of the gauge generators in 
phase space: They are constructed with linear combinations of 
constraints with arbitrary functions and their first time derivatives.  
Let us consider two of these generators $G[\xi_1], G[\xi_2]$.  Their 
Poisson Bracket, an equal time commutator, is on general grounds 
$G[\xi_3]$ for some $\xi_3$.  It is impossible to get for $\xi_3$ the 
standard diffeomorphism rule
$\xi^\mu_3 = \xi_2^\nu \xi_{1,\nu}^\mu - \xi_1^\nu \xi_{2,\nu}^\mu $: 
In such a case $\dot\xi_3$, which appears in $G[\xi_3]$, will depend 
on the second time derivatives of $\xi_1$ and $\xi_2$, and this 
dependence {\it cannot} be generated by the equal time Poisson Bracket 
$\{G[\xi_1], G[\xi_2]\}$.  Nesting of Poisson Brackets would introduce 
yet higher time derivatives.  This is why general reparameterization 
covariance cannot be implemented in this form in the Hamiltonian 
formalism.  The argument applies to any reparameterization covariant 
theory.

This was the argument used in \cite{Salisbury83} to realize 
diffeomorphisms in the canonical formalism.  In fact the arena in 
\cite{Salisbury83} was the reduced space defined by the variables 
$(g_{a b}, K^{a b})$.  In this case there are no time derivatives of 
the arbitrary functions $\xi^{\mu}$ in the variations generated by 
(\ref{thegen}), but the argument still applies in the same way, as 
shown above.  Once the obstruction to projectability is identified 
through the form of $[\delta_1,\delta_2] x^\mu $, the assumption of a 
metric dependence in $\epsilon^\mu$ and the requirement that 
$[\delta_1,\delta_2] x^\mu $ must not have any explicit time 
derivative of $\epsilon_1$ or $\epsilon_2$ leads to equations 
(\ref{3.10}-\ref{3.13}), the projectability condition.

We should caution that the algebra (\ref{5.20}-\ref{5.22}) is 
satisfied only under the condition that there is no other gauge 
symmetry in addition to diffeomorphism-induced symmetry.  Under more 
general circumstances, pure diffeomorphisms (even the field dependent 
variety given by \ref{3.14}) are not realizable as canonical 
transformations; they must be accompanied by related internal gauge 
transformations \cite{Salisbury83b}.  The issue of projectability for 
these models will be addressed in another paper.

Gauge theories like electromagnetism or Yang-Mills in Minkowski 
spacetime share with general relativity the property that gauge 
transformations (diffeomorphisms in general relativity) need to be 
constructed with arbitrary functions and their spacetime first 
derivatives.  The gauge generators are made up of two pieces, 
associated with a primary and a secondary constraint; it is therefore 
mandatory that all these theories have secondary constraints.  This is 
the way by which the canonical formalism is able to provide us with 
the right gauge transformations.

For the sake of completeness, let us now apply these ideas to our 
relativistic particle (\ref{partlag}).  The gauge generator is, from 
(\ref{thegen}), 
$$ 
    G(t) = {\dot \xi}(t) \pi + \xi(t) \frac{1}{2} (p_\mu p^\mu + 1), 
$$ 
with $\xi$ an arbitrary function of time.  One can easily check that 
$G(t)$ generates the (projectable) transformations (\ref{del-x}) 
introduced above.  Notice also that if $\xi$ is a constant, the 
secondary constraint generates a rigid (time-independent) Noether 
symmetry, whereas the primary one does not.  Primary constraints 
generate gauge symmetries only in the case when they do not lead to 
secondary constraints through the stabilization algorithm.

Finally, notice that we do not modify ``by hand'' the 
Hamiltonian by adding to it the secondary constraint with a new 
Lagrange multiplier.  This modification, the so called Dirac 
conjecture, turns out not only to be unnecessary but to break the 
equivalence with the Lagrangian theory as well \cite{Gracia88}.


\section{Gauge Fixing and Reduced Formalism}

\subsection{The Gauge Fixing Procedure}

One of the methods to eliminate the superfluous degrees of freedom of 
a gauge theory is through the introduction of a new set of 
constraints.  This is the gauge fixing procedure, which according to 
\cite{Pons95}, can be performed in two different steps: the first is 
to fix the dynamics, the second to fix the redundancy of the initial 
conditions (though this need not be the order in which the whole set 
of constraints is introduced).

First, to fix the dynamics---to determine specific values for the 
functions $\lambda^\mu$ in (\ref{dirach})---we must introduce $M+1$ 
constraints, $\varphi_\mu \simeq 0 $, such that 
$\det|\{\varphi_\mu , P_\nu \}| \neq 0$.  A typical set could be 
\begin{equation} 
    \varphi_\mu = N^\mu - f^\mu,
    \label{gaugef}
\end{equation} 
with $f^\mu$ ($f^0 \neq 0$) a given set of functions not depending on 
$P_{\mu}$ or $N^{\nu}$ (the simplest choice could be 
$f^a = 0,\, f^0 = 1$).  We could also think of $f^\mu$ 
as a not yet determined set of functions.  These gauge fixing 
constraints fix $\lambda^\mu$ in $H_{\rm D}$ to be zero and then:
\begin{equation} 
    H_{\rm D}^{\small red} 
    = N^\mu {\cal H}_\mu 
    \cong f^\mu {\cal H}_\mu,  
        \label{hred} 
\end{equation} 
where we have used Dirac's notation of strong equality, $\cong$, to 
mean an equality up to {\it quadratic} pieces in the constraints, 
including the gauge fixing ones.  In practice this strong equality 
tells us that we can substitute $f^\mu$ for $N^\mu$ within the 
Hamiltonian due to the fact that ${\cal H}_\mu$ are constraints, too.

Once this set of evolution-fixing constraints $\varphi_\mu \simeq 0 $ 
has been introduced, with a given set of functions $f^\mu$, the gauge 
transformations---strictly speaking---have disappeared.  In fact, if 
we require the gauge generators to be consistent with the new 
constraints, $\{\varphi_\mu ,\, G(t) \} = 0$, we get the relations 
\begin{equation} 
    \dot \xi^\mu + \xi^\nu N^\sigma C^\mu_{\nu\sigma} = 0, 
    \label{nogaug} 
\end{equation} 
which means that the functions $\xi^\mu$ cease to be arbitrary (at 
least with respect to the time dependence), and hence there are no 
more gauge transformations.  The transformations $G(t)$ satisfying 
(\ref{nogaug}) can be called, as is usual in other contexts, residual 
gauge transformations, but it must be emphasized that they are not 
true gauge transformations in phase space, because the arbitrariness 
that was present in the Dirac Hamiltonian has been eliminated.  We 
encounter a parallel case in electromagnetism, for instance, when 
after introducing the Lorentz gauge $\partial_\mu A^\mu = 0$, we are 
left with a residual gauge symmetry, 
$A^\mu \rightarrow A^\mu + \partial_\mu \Lambda$, 
provided $\Lambda$ satisfies $\Box \Lambda = 0$.

Another way to view the residual gauge transformations, which is more 
interesting to us, is to consider the situation at a given initial 
time, $t=0$.  Let 
$\alpha^\mu({\bf x}) = \xi^\mu(0,{\bf x}),\, 
\beta^\mu({\bf x}) = \dot\xi^\mu(0,{\bf x})$; 
they are related by (\ref{nogaug}): 
$\beta^\mu + \alpha^\nu N^\sigma C^\mu_{\nu\sigma} = 0$.  
We are left with the ``residual'' gauge transformation at $t=0$
$$ 
    G_{\rm R}(0) = P_\mu \beta^\mu 
     + ( {\cal H}_\mu + N^\rho C^\nu_{\mu\rho} P_\nu) \alpha^\mu 
    = {\cal H}_\mu \alpha^\mu , 
$$
with $\alpha^\mu$ an arbitrary function of $M$-space variables.

The role of $G_{\rm R}(0)$ is that it generates transformations on the 
initial value surface that describe a redundancy that is still left in 
the formalism, and we must eliminate it in order to arrive at the true 
degrees of freedom.  Thus we must introduce a new set of gauge fixing 
constraints, $\chi_\mu \simeq 0$, with the requirements:
\begin{enumerate}
\item
The dynamical evolution, which is already fixed, must preserve these 
new constraints.
\item
$\{ \chi_\mu , G_{\rm R}(0) \} = 0 $ must imply $\alpha^\mu = 0$.
\end{enumerate}
Obviously, to satisfy the second condition, we need 
$\det|\{\chi_\mu , {\cal H}_\nu \}| \neq 0$, 
and to satisfy the first we need: 
\begin{equation} 
    \{\chi_\mu, H_{\rm D} \} + {\partial \chi_\mu\over\partial t} 
    = f^\nu \{\chi_\mu, {\cal H}_\nu \} 
        + {\partial\chi_\mu\over\partial t} \simeq 0.  
\end{equation}
Notice that the first and the second conditions are only compatible if 
at least one of the gauge fixing constraints $\chi_\mu$, for instance 
$\chi_0$, has explicit time dependence: $\partial{\chi_0}/\partial{t} 
\neq 0$.  This result implies that time needs to be defined 
classically through a function of the canonical variables.


\subsection{The Reduced Formalism}

Notice that if we perform the partial gauge fixing defined in 
(\ref{gaugef}), $N^\mu - f^\mu \simeq 0$ in the spirit of keeping 
$f^\mu$ undetermined, then we can interpret
$$  
    H_{\rm D}^{\small red} = f^\mu {\cal H}_\mu 
$$ 
in (\ref{hred}) as the Hamiltonian for the reduced phase space 
described by all variables other than $P_{\mu}$ and $N^{\nu}$.  In 
this reduced space we have a dynamical theory defined by a vanishing 
canonical Hamiltonian and a set of constraints, which now become 
primary, ${\cal H}_\mu \simeq 0$.  Then the new Dirac Hamiltonian is 
$H_{\rm D}^{\small red}$ and the new gauge generator is
$$ 
    G^{\small red} = \xi^\mu {\cal H}_\mu.  
$$ 
Thus we see that the constraints ${\cal H}_\mu$ do generate gauge 
transformations {\it in the reduced phase space}.

We identify here a frequent source of confusion in the literature when 
it is claimed that all first class constraints, either primary or 
secondary (or tertiary, etc.), generate gauge transformations.  For 
generally covariant theories with a metric, in the original phase 
space, only specific combinations, as in (\ref{thegen}), of primary 
and secondary constraints generate gauge transformations.  But in the 
reduced formalism, since the old secondary constraints take the role 
of primary constraints and there are no more constraints, these new 
primary constraints generate gauge transformations in the reduced 
space.

As to the gauge fixing procedure in the reduced phase space, since 
there are only primary constraints, there is only one step to be 
undertaken: to fix the evolution.  Notice that the same argument we 
used previously to show that one of the gauge-fixing constraints must 
be the definition of time applies here as well.


\subsection{From the Reduced to the Original Formalism}

In the case of generally covariant theories, we have seen that the 
reduced formalism consists in the elimination of the primary 
constraints, $P_\mu$, and their canonical conjugate variables, 
$N^\mu$, through a partial gauge fixing 
$N^\mu = f^\mu, \, \, (f^0 \neq 0)$, with $f^\mu$ 
arbitrary functions of spacetime as well as of 
the reduced variables.  Then we obtain a reduced 
theory which has ${\cal H}_\mu \simeq 0$ as primary constraints (no 
secondary constraints appear), 
$H^{\small red} = f^\mu {\cal H}_\mu$ 
as the Dirac Hamiltonian, and 
$G^{\small red} = \xi^\mu {\cal H}_\mu$ 
as the generator of gauge symmetries.  The new 
bracket for the set of the reduced variables is just the Dirac 
Bracket, which in our case is trivially obtained as the old Poisson 
Bracket when acting with the reduced variables.

One may wonder whether there is a way to restore the full theory from 
the reduced one.  In these cases where the constraints eliminated in 
the process of reduction are canonical momenta we will see that there 
exists such a method.  This is the enlargement procedure:

Consider a theory defined by a canonical Hamiltonian $H_{\rm C}$ and a 
set of primary constraints $\phi_\mu \simeq 0$.  The Dirac Hamiltonian 
is $H_{\rm D} = H_{\rm C} + \lambda^\mu\phi_\mu$ ($\lambda^\mu$ are 
arbitrary functions).  Let us suppose we have applied the 
stabilization algorithm to obtain secondary, tertiary, etc., 
constraints and that we have finally obtained a set of Noether gauge 
generators, described by a single $G[\xi]$, where $\xi$ stands for a 
set $\xi^\alpha$ of infinitesimal arbitrary functions of spacetime.  
$G[\xi]$ is assumed to be a local functional of $\xi^\alpha$ (that is, 
it depends linearly on $\xi^\alpha$ and a finite number of its time 
derivatives, according to the length of the stabilization algorithm).  
We also assume the commutation algebra for $G[\xi]$ to be
$$ 
    \{G[\xi_1] , \, G[\xi_2] \} = G[\xi_3],
$$
with 
$\xi_3^\alpha = C_{\beta\gamma}^\alpha \xi^{\beta}_1 {\xi}^{\gamma}_2$ 
(this is a general property, and the $C_{\beta\gamma}^\alpha$ are not 
necessarily the same as previously defined; remember that repeated 
indices imply both summation and integration).

According to Section 2, there exist functionals $A^\mu_\nu[\xi]$ and 
$B^\mu[\xi]$ such that
$$  
    \{G[\xi],H_{\rm C}\} + {\partial G[\xi]\over\partial t} 
        = B^\mu[\xi]\phi_\mu, \quad 
    \{G[\xi],\phi_\nu \} 
        = A^{\mu}_\nu[\xi]\phi_{\mu}.  
$$ 

The enlargement procedure consists in promoting the arbitrary 
functions $\lambda^\mu$ to the status of canonical variables; let us 
call them $N^\mu$ for obvious reasons.  Let us introduce canonical 
momenta $P_\mu$ associated with the new variables (we thus trivially 
enlarge the Poisson Bracket) and require these new momenta to be the 
primary constraints of the enlarged theory.

The enlarged canonical Hamiltonian will then be 
$H_{\rm C} + N^\mu\phi_\mu$, and the new Dirac Hamiltonian will be 
$H_{\rm E} = H_{\rm C} + N^\mu\phi_\mu + \eta^\mu P_\mu$, 
with $\eta^\mu$ new arbitrary functions.  It is straightforward to 
verify that the dynamics of the original theory and that of the 
enlarged theory coincide as far as the evolution of the original 
variables is concerned.  However, note that the $\phi^\mu$ have become 
secondary constraints.

Now we will show how to enlarge the corresponding gauge generators 
$G[\xi]$.  Since the new primary constraints $P_\mu$ must appear 
within the enlarged gauge generators $G_{\rm E}[\xi]$, we will assume 
the general form $G_{\rm E}[\xi] = G[\xi] + S^\mu[\xi] P_\mu$, with 
$S^\mu$ to be determined through the requirements of Section 2.  It 
turns out that $S^\mu = B^\mu + N^\nu A^\mu_\nu P_\mu$.  The enlarged 
gauge generator therefore has the following form: 
\begin{equation} 
    G_{\rm E}[\xi] = G[\xi] + B^\mu[\xi] P_\mu 
          + N^{\nu} A^{\mu}_{\nu}[\xi] P_{\mu}.  \label{enlargen} 
\end{equation} 
The commutation algebra for $G_{\rm E}[\xi]$ is: 
$$ 
    \{G_{\rm E}[\xi_1] , \, G_{\rm E}[\xi_2] \} 
    = G_{\rm E}[\xi_3] + {\cal O}(P^2), 
$$ 
where ${\cal O}$ is pure quadratic in the (new) primary constraints:
\begin{eqnarray}
  &&{\cal O}(P^2) \nonumber\\
   &&~~     = \left\{B^\mu[\xi_1] + N^\nu A^\mu_\nu[\xi_1] , \, 
            B^\sigma[\xi_2] + N^\rho A^\sigma_\rho[\xi_2] \right\} 
                P_\mu P_\sigma.\nonumber 
\end{eqnarray}  
In our particular case of general covariant theories, 
$B^\mu[\xi] = \dot\xi^\mu$ and 
$A^\mu_\nu[\xi] = \xi^\rho C_{\rho\nu}^\mu$.  Notice 
that in this particular case the term ${\cal O}(P^2)$ vanishes.

The procedure of enlargement here devised is completely general, and 
it is valid for any gauge theory no matter how complicated its 
structure of constraints may be.


\section{The Gauge Group}

The gauge group is a subgroup of the symmetry group of the system.  A 
symmetry is a transformation that maps solutions of the equations of 
motion into solutions.  From a physical standpoint, gauge symmetry 
reflects a redundancy in the description.  Mathematically, a gauge 
transformation is characterized by its functional dependence on 
arbitrary functions.  The functional dependence is expected to be 
local in the sense of depending on the values of the functions and on 
a finite number of derivatives.  This is the definition for classical 
mechanics and classical field theory.  It is most convenient to define 
gauge symmetries in a more restrictive way as local transformations 
which leave the action invariant up to boundary terms.  Our analysis 
is based on the Noether identities which result from this invariance 
under infinitesimal local symmetries.

Let us make some formal remarks on the nature of the 
diffeomorphism-induced gauge group of the type discussed in this 
paper.  Let ${\it Riem}({\cal M})$ be the space 
of (pseudo) Riemannian metrics of the spacetime manifold ${\cal M}$, 
and let ${\bf Diff}({\cal M})$ be the group of diffeomorphisms in 
${\cal M}$.  An element of the gauge group 
${\bf G}[{\it Riem}({\cal M})]$ is a regular map 
${\it Riem}({\cal M}) \rightarrow {\it Riem}({\cal M})$ 
such that each $g \in {\it Riem}({\cal M})$ 
undergoes a diffeomorphic transformation, that is, a transformation 
dictated by a specific element of ${\bf Diff}({\cal M})$ 
(thus keeping the action invariant).  (Other fields are also affected 
by this diffeomorphic transformation, but for this discussion we 
devote our attention to the metric.)  But this element of 
${\bf Diff}({\cal M})$ may be different if we consider 
the action of the same element of the gauge 
group on a different $g' \in {\it Riem}({\cal M})$.

To determine an element of the gauge group we must assign to each 
$g \in {\it Riem}({\cal M})$ the specific spacetime 
diffeomorphism which is going to act on $g$.  More precisely, 
an element, $d$, of the gauge group is a map 
\begin{eqnarray} \nonumber  
    d: {\it Riem}({\cal M}) &\longrightarrow& 
                        {\bf Diff}({\cal M})\\
                 g &\longrightarrow& d[g] 
\end{eqnarray} 
such that we can build out of it a regular map 
\begin{eqnarray} \nonumber
      {\bf G}: {\it Riem}({\cal M}) &\longrightarrow& 
              {\it Riem}({\cal M})\\ 
        g &\longrightarrow& (d[g])(g) .
\end{eqnarray}

Now let us consider the generators of the gauge group {\bf G}.  We 
use, for the sake of generality, a condensed notation where $\Phi^i$ 
stands for the fields that are present in the theory; the action is 
denoted by ${\cal S}$, and $i$ includes continuous spacetime indices 
(so that repeated indices imply both summation and integration).  Let 
$\epsilon^\alpha$ be arbitrary functions of spacetime variables, and 
$\delta_\epsilon \Phi^i = R_\alpha^i \epsilon^\alpha$ be a complete 
set of infinitesimal gauge transformations.  These satisfy the Noether 
identities $(\delta {\cal S} / \delta \Phi^i) R_\alpha^i = 0$.  
Obviously we do not alter the Noether identities by taking a different 
linear combination of variations 
$R_\alpha^i \rightarrow \bar R_\alpha^i 
= \Lambda_\alpha^\beta(\Phi) R_\beta^i$, even when the 
$\Lambda_\alpha^\beta$ depend on $\Phi^i$.  No gauge equivalent 
trajectories are eliminated through this transformation, presuming 
that $\bf\Lambda$ is invertible.  In our case (see \ref{3.14}) the 
requirement of projectability fixed
\begin{equation} 
    \Lambda^\mu{}_\nu (g) 
        = n^\mu \delta^0_\nu +\delta^\mu_a \delta^a_\nu, 
\end{equation} 
which is clearly invertible.  The algebra corresponding to this new 
choice of generators contains field-dependent structure coefficients.

One might conclude that this ``soft'' algebra structure signifies that 
the symmetry transformations no longer form a group.  We do have a 
group, which acts not on the spacetime manifold $\cal M$ but on ${\it 
Riem}({\cal M})$.  Note that elements of the diffeomorphism-induced 
gauge group must depend on the full metric.  Dependence on the lapse 
and shift is fixed by (\ref{3.14}).  Perhaps more surprising is the 
fact that the full group depends non-locally on the hypersurface 
induced metric.  This is a direct consequence of the structure 
coefficients in (\ref{5.22a}): Repeated nesting of the commutator 
produces spatial derivatives of $g_{a b}$ to infinite order 
\cite{Bergmann72}.

To summarize, we have discussed some aspects of the canonical approach 
to generally covariant theories.  In particular we have emphasized the 
special way in which the canonical formalism describes the 
diffeomorphism covariance of these theories.  The gauge group for 
these theories is larger than the diffeomorphism group.  The canonical 
gauge generators are just one of the possible bases for the gauge 
algebra, although projectability of transformations generated by the 
larger group from configuration-velocity space to phase space fixes 
the dependence on the lapse and shift uniquely.  We have displayed the 
canonical generators for the gauge symmetries of these theories on the 
entire phase space.  Transformations may be pulled back to the entire 
configuration-velocity space.

In this paper we assumed that the only gauge symmetries are generated 
by diffeomorphisms.  When other gauge symmetries occur, related 
internal gauge transformations must be taken into account.  This topic 
and the question of projectability of the gauge transformation group 
onto the full constraint hypersurface will be dealt with in future 
papers.  It is also our intention to explore further the relationship 
between our results and matters pertaining to quantization, 
particularly the question of time in quantum gravity.


\section*{Acknowledgments}

JMP and DCS would like to thank the Center for Relativity of The 
University of Texas at Austin for its hospitality.  JMP acknowledges 
support by CIRIT and by CICYT contracts AEN95-0590 and GRQ 93-1047.  
DCS acknowledges support by National Science Foundation Grant 
PHY94-13063.


\appendix\section*{The Nambu-Goto Relativistic String}

We have so far discussed the situation in which the metric may be 
constructed using only configuration space variables.  Our 
conclusions, however, do hold in case that velocity variables, too, 
are needed to construct the metric.  In this Appendix we illustrate 
the issue of projectability of variations engendered by 
diffeomorphisms in the Nambu-Goto relativistic string.  This is an 
example of a first order dynamical model in which a metric may be 
constructed using velocity as well as configuration variables.  (The 
Polyakov string satisfies the conditions postulated in Section 3.)

We let $\{y^I,\ I=0,\dots,M\}$ represent Minkowski spacetime 
coordinates.  The string surface is given by $y^I(x^\mu),\ \mu=0,1$, 
where $x^0 = \tau$ and $x^1 = \sigma$.  The induced metric on the 
string surface is 
\begin{equation} 
    g_{\mu \nu} = y^I_{,\mu}y^J_{,\nu}\; \eta_{I J}.  
    \label{str.111} 
\end{equation} 
For spacetime contractions we use the notation 
($~\dot{}=d/d\tau, {}'=d/d\sigma$): 
\begin{equation}
    (\dot y^2)=\dot y^I\dot y_I,\ 
    (y'^2)=y'^Iy'_I,\ 
    (\dot yy')=\dot y^Iy'_I. 
\end{equation} 
The Lagrangian density is minus the string volume element : 
\begin{equation} 
    L=-(-\det g)^{1/2}
    =-\big(-(\dot y^2) (y'^2) +(\dot yy')^2\big)^{1/2}.  
    \label{str.222} 
\end{equation} 
From 
\begin{equation} 
    g^{\mu \nu} = -L^{-2} 
    \pmatrix{ (y'^2) &-(\dot yy')\cr -(\dot yy')&(\dot y^2)}, 
        \label{str.333} 
\end{equation} 
we read off the lapse and shift: 
\begin{equation}
    N=\frac{L}{(y'^2)^{1/2}}, 
    \label{str.444} 
\end{equation}
and 
\begin{equation} 
    N^1 = \frac{(\dot yy')}{(y'^2)}.  
    \label{str.555}
\end{equation} 
The canonical momentum is 
\begin{equation} 
    {\hat \pi}^I={\partial L \over \partial {\dot y}_I} 
    = -L^{-1} \big( (y')^2 \dot y^I - (\dot yy') y'^I).
        \label{str.666} 
\end{equation}

Recall that when we are working in configuration-velocity space, the 
coordinates are $\{y^I,\dot y^I\}$.  There are two primary 
constraints in phase space:
\begin{equation} 
    \phi_0 =\frac{1}{2}\big((\pi)^2 + (y')^2 \big), 
    \label{str.777} 
\end{equation} 
and 
\begin{equation} 
    \phi_1 = (y'\pi). \label{str.888} 
\end{equation} 
Therefore one may ask whether the velocities may be expressed uniquely 
in terms of the canonical momenta and the lapse and shift, taking into 
account that these constraints show that there are a correct number of 
coordinates $\{y^I,\hat\pi^I,N,N^1\}$ for velocity space.  In our 
example we can indeed invert the expression for $\hat\pi^i$ to obtain
\begin{equation} 
    {\dot y}^I = N^1 y'^I - \frac{N}{(y'^2)^{1/2}} \hat\pi^I. 
    \label{str.999} 
\end{equation} 
The primary constraints are relations among the $y^I$ and the 
$\pi^I$ and involve neither lapse nor shift.  Therefore invertability 
is equivalent to the demand that 
\begin{equation} 
    0={\partial {\hat \pi}_I \over \partial N^\mu} 
    = {\partial {\hat \pi}_I \over \partial {\dot y}^J} 
        {\partial {\dot y}^J \over \partial N^\mu} 
    = W_{I J} {\partial {\dot y}^J \over \partial N^\mu},
                 \label{str.1111} 
\end{equation}
where 
\begin{eqnarray} 
    W_{IJ} &=& {\partial^2L \over \partial \dot y^I 
        \partial \dot y^J} 
        \nonumber\\
    &=& -\frac{1}{L^3}\big( L^2 (y'^2) \delta_{IJ} 
        +(\dot y^2)(y'^2) y'_I y'_J +(y'^2)^2 \dot y_I 
            \dot y_J
            \nonumber\\ 
        &&~~~~~~-(\dot y y')(y'^2) (\dot y_I y'_J + y'_I 
            \dot y_J)\big) 
            \nonumber\\ 
    &=& -\frac{1}{N(y'^2)^{1/2}} \left((y'^2)\delta_{IJ} 
        -y'_Iy'_J +\hat\pi_I\hat\pi_J\right)\ .  
\end{eqnarray} 
But this is the statement that
$$  
    \frac{\partial\dot y^I}{\partial N} 
    =-\frac{1}{(y'^2)^{1/2}}\hat\pi^I\ {\rm and}\ 
    \frac{\partial\dot y^I}{\partial N^1}=y'^I 
$$ 
are null eigenvectors of $W_{IJ}$; that they are is readily verified.

Recall that in our model the primary constraints are 
$ \phi_0 =\frac{1}{2}\left((y')^2 + (\pi^2) \right)$ 
and $ \phi_1 = (y'\pi)$.  We have 
\begin{equation} 
    \gamma^I_0 = {\cal F}\!L^* 
        \left( {\partial \phi_0 \over \partial \pi_I} \right) 
        = {\hat \pi}^I \ {\rm and}\ 
    \gamma^I_1 = {\cal F}\!L^* 
        \left( {\partial \phi_1 \over \partial \pi_I} \right) 
        = y'^I. 
            \label{str.2222} 
\end{equation} 
Thus 
\begin{equation} 
    {\bf\Gamma}_0 = {\hat \pi}^I {\partial \over \partial {\dot y}^I}
        = -(y'^2)^{1/2} {\partial \over \partial N} \ {\rm and}\ 
    {\bf\Gamma}_1 = y'^I{\partial \over \partial {\dot y}^I}
        = {\partial \over \partial N^1}. 
            \label{str.3333} 
\end{equation} 
Therefore variations of the $y^I$ will be projectable 
if and only if they are independent of $N$ and $N^1$.  These 
variations are 
\begin{eqnarray} 
    \delta y^I &=& y^I_{,\mu}\epsilon^\mu 
    = \dot y^I\epsilon^0 +y'^I\epsilon^1 \nonumber\\
    &=& \left(N^1 y'^I 
        - \frac{N}{(y'^2)^{1/2}} \hat\pi^I\right) \epsilon^0 
        + y'^I \epsilon^1.  
            \label{str.4444} 
\end{eqnarray} 
It is straightforward to show that these 
variations will be independent of $N,N^1$ if and only if 
\begin{equation} 
    \epsilon^0 = \frac{1}{N}\xi^0,\ \epsilon^1 
    = -\frac{N^1}{N}\xi^0 +\xi^1, 
        \label{str.5555} 
\end{equation} 
where $\xi^\mu$ are independent of $N,N^1$.  In terms of the 
timelike unit vector 
$$  
    n^\mu=(1/N,-N^1/N),
$$
this result is in the correct form, namely the same as (3.13): 
\begin{equation} 
    \epsilon^\mu = \delta^\mu_1 \xi^1 + n^\mu \xi^0.  
    \label{str.6666} 
\end{equation}


\end{multicols}

\end{document}